\documentclass[12pt]{article}
\usepackage{}
\begin{document}
\date{}
\title{\textbf{Crypto-Harmonic Oscillator in Higher Dimensions: Classical and Quantum Aspects}}
\author{ {Subir Ghosh}\thanks{E-mail:subir\_ghosh2@rediffmail.com}
\\\textit{Physics and Applied Mathematics Unit,}
\\\textit{Indian Statistical Institute, 203 B. T. Road, Kolkata 700108, India}\\
\\
 and\\ \\{Bibhas Ranjan Majhi}\thanks{E-mail: bibhas@bose.res.in}
\\\textit{S.~N.~Bose National Centre for Basic Sciences,}
\\\textit{JD Block, Sector III, Salt Lake, Kolkata-700098, India}}
\maketitle

\begin{quotation}
\noindent \normalsize
\end{quotation}
Abstract:\\
We study complexified Harmonic Oscillator models in two and three
dimensions. Our work is a generalization of the work of Smilga
\cite{sm} who initiated the study of these Crypto-gauge invariant
models that can be related to $PT$-symmetric models. We show that
rotational symmetry in  higher spatial dimensions naturally
introduces  more  constraints, (in contrast to \cite{sm} where one
deals with a single constraint), with a much richer constraint
structure. Some common as well as distinct features in the study
of the same Crypto-oscillator in different dimensions are
revealed. We also quantize the two dimensional Crypto-oscillator.
\vskip .5cm {\bf{Introduction:}} It has been known for quite
sometime \cite{early} that there are quantum mechanical models
with specific complex terms in the Hamiltonian that admit real
spectra and unitary evolution. Later the seminal paper of Bender
and Boettcher \cite{ben}  attributed this intriguing and useful
property to the combined $PT$ (parity and time reversal) symmetry
of the system and more $PT$-symmetric models were constructed that
had the above feature. Subsequently there has been a lot of
activity \cite{others} in the study of different aspects of
$PT$-symmetric models. These models are referred as
``Crypto"-Hermitian models by Smilga \cite{sm}. In \cite{sm}
Smilga has also provided an alternative explanation to this
behavior (of having real energy eigenvalues for a complex
Hamiltonian): Crypto-gauge invariance. However, in an important earlier work by Mostafazadeh \cite{ali}, it was observed in a general context that the real part of the Hamiltonian can generate the dynamics in a real phase space and that the imaginary part of the Hamiltonian, treated as a constraint, can generate symmetry transformations. The usage of certain class of coordinates in previous works \cite{early1} in related problems was also explained in \cite{ali}. The idea is to complexify a
real Hamiltonian system and subsequently treat the real part of
the complex ${\cal{H}}$ as the Hamiltonian $H$ of the enlarged
system with twice the original number of degrees of freedom. By
virtue of Cauchy-Riemann condition (for ${\cal{H}}$) and
Hamiltonian equations of motion it is possible to show that both
the real part $H$ and the imaginary part $G$ of ${\cal{H}}$ (where
${\cal{H}}=H+iG$), are {\it{separately conserved}}. This allows
one to interpret $G$ as a First Class Constraint (FCC) (see section II for a brief discussion on the constraint analysis as formulated by Dirac \cite{dir})
and in particular $G=0$ ensures reality of the energy value. This
FCC is present in all such complexified systems and the gauge
symmetry induced by this FCC \cite{dir} is termed as Crypto-gauge
symmetry \cite{sm}. In \cite{sm} it has been shown that specific
features of some complexified models, (analyzed in terms of real
variables), can be matched with their $PT$-symmetric counterpart
in the complex plane.

It is important to emphasize that the work of Smilga \cite{sm} is
restricted to one space dimension only and it naturally evokes the
question of its application in higher dimensions. The present work
specifically deals with this problem where we study the
complexified or "Crypto" Harmonic Oscillator (CHO) in two and
three dimensions. The one dimensional Crypto-oscillator was
discussed by Smilga in \cite{sm}. As we will discuss at length in
this paper, even this straightforward generalization reveals a
number of interesting features that demand further study of higher
dimensional Crypto-gauge systems in a model independent way. It is
worth mentioning that not much work has been done in
$PT$-symmetric models in higher dimensions. Indeed,  it will be
very fruitful if, our way of studying Crypto-gauge invariant
models can reproduce results that our comparable with previously
studied higher dimensional $PT$-symmetric models \cite{lev}.

In this article we will concentrate on additional spatial
symmetries, (such as rotational symmetries), that naturally occur
in more than one dimensions. Following the same philosophy of
demanding reality of energy values, which is a conserved quantity,
one can also demand reality of {\it{other conserved quantities}},
such as angular momentum, (as we have done here). This induces
more constraints in the system and the subsequent analysis will
require the Hamiltonian constraint analysis \cite{dir}. Our study will reveal  a rich and interesting
constraint structure for the higher dimensional models.

There seems to be still another way of interpreting the appearance
of Crypto-gauge symmetry in this complexification process. In
quantum field theories in the area of High Energy Physics, there
are several systematic procedures \cite{bat} of introducing gauge
invariance (by way of FCCs) where the original model is embedded
in a prescribed way in an extended phase space. The equivalence of
the extended gauge invariant model with the original model is
established in the so called unitary gauge where the extended
model reduces to the original one. Here it is essential for the
extended model to have the requisite number of FCCs that can
account for the additional degrees of freedom in the enlarged
phase space.

It is quite intriguing that the same phenomenon is repeated in the
Crypto-gauge symmetric models although this was not quite apparent
in the one dimensional examples studied in \cite{sm}. In one
dimension, complexification introduces one extra degree of freedom
and there appears the FCC $G\approx 0$ to remove it. In fact
examples of unitary gauge choices have been given in \cite{sm}. On
the other hand, in the higher dimensions that we consider, (albeit
in the CHO model), larger number of degrees of freedom are
introduced in the complexification process but quite surprisingly
the number and nature of the additional constraints that appear
from other conserved quantities (such as angular momentum) are
just right to account for the extra variables. Indeed it will be
very interesting to establish this property in
higher dimensions in a model independent way.\\
\vskip .5cm {(II) \bf{Dirac Constraint Analysis - A Brief Digression:}}
In the coordinate space formulation, starting from a Lagrangian
$L(q_i,\dot q_i)$ of a dynamical system, constraints (if present) are revealed
from the definition of the canonically conjugate momenta
$p_i=(\partial L)/(\partial \dot q_i)$. In the Hamiltonian scheme \cite{dir},  constraints are a set of relations  $\varphi_a(q_i,p_i)\approx 0$, without any time derivative. The weak equality stresses the fact that the constraints can be put to zero $\varphi _a=0$ as a strong equality only after all the relevant Poisson brackets are computed. Other constraints can
appear from the requirement that the constraints are preserved in
time and a complete set of constraints should obey 
$$\{\varphi _a,H\}=0 +\lambda _a\varphi _a.$$
 Here
$\varphi _a$ are a set of independent constraints and
$$H(p_i,q_i)=p_j\dot q_j-L +\lambda _a\varphi _a$$ is the
canonical Hamiltonian modulo constraints. The Poisson bracket is
computed by using the basic algebra
$$\{q_i,p_j\}=\delta_{ij}~,\{q_i,q_j\}=\{p_i,p_j\}=0.$$ 

Once the
full set of constraints are obtained Dirac introduced the very
important classification of constraints. If in the full set
$\varphi _a$, there are constraints $F_\alpha$ that (Poisson)
commute with {\it{all}} the constraints,
$$\{F_\alpha ,\varphi _a \}=0 +\lambda_{\alpha ab}\varphi _b,$$
 the set $F_\alpha$ are
termed as First Class Constraints (FCC). The rest of the
constraints $H_\beta $ that do not commute with all the
constraints are termed as  Second Class Constraints (SCC). In
practical terms this means that the constraint matrix, with $\{\varphi _a,\varphi _b\}$ as matrix elements, will be
{\it{degenerate}} if there are FCCs in the system and it will be
invertible if only SCCs are present.

The FCCs are responsible for
local gauge invariances in the system and they are related to the
generators of local gauge transformations. On the other hand, the
SCCs induce a modification in the symplectic structure and one has
to replace the basic Poisson Brackets by a new set of brackets,
known as Dirac Brackets. Also it is important to point out that the
presence of FCCs indicate that there are redundant variables that
are not physical degrees of freedom and one is allowed to choose
additional constraints, known as gauge fixing conditions, that can remove
these trivial variables. Notice that a system of FCCs together
with proper gauge fixing constraints becomes a set of SCCs.

An SCC can be used to eliminate one degree of freedom in phase space. On the other hand, one FCC, together with an associated  gauge fixing constraint, constitute a pair of SCCs and  accounts for two degrees of freedom in phase space. In this way one can determine the true degrees of freedom of a constrained system.

The
idea is that in quantizing a system with Second Class Constraints,
one needs to elevate the Dirac Brackets, (and {\it{not}} the Poisson Brackets), to quantum commutators.

In the present work we will only invoke the idea of classification
of constraints and explicit construction of the Dirac Brackets
will be left for a future publication. 
\vskip .5cm
{(III) \bf{$2$-Dimensional CHO: Classical Analysis:}}  The CHO
Hamiltonian is,
\begin{eqnarray}
{\cal{H}}(\pi_i,z_i) = \frac{{\pi_i}^2+{z_i}^2}{2} \label{1.1}
\end{eqnarray}
where $i=1,2$. Clearly this is just the two-dimensional extension
of the  construction of Smilga \cite{sm}. Following \cite{sm} we
express  the complex phase space variables $(\pi_{i};z_{i})$ in
terms of real phase space variables $ z_i =
x_i+iy_i~,~\pi_i=p_i-iq_i $. The above phase space is canonical
with the only non-vanishing Poisson brackets being $
\{x_i,p_j\}=\delta_{ij}~;~ \{y_i,q_j\}=\delta_{ij}$. The  complex
Hamiltonian  ${\cal {H}}$ in (\ref{1.1}) now reads,
$$
{\cal{H}}(\pi_i,z_i) = H(p_i,q_i;x_i,y_i)+iG(p_i,q_i;x_i,y_i),$$
\begin{equation}
H=\frac{1}{2}[({p_i}^2+{x_i}^2)-({q_i}^2+{y_i}^2)],~ G=-p_i q_i +
x_i y_i . \label{1.5}
\end{equation}
In order to restrict the classical Hamiltonian  to the real space,
we impose the constraint $G \approx 0$ where the weak equality is
interpreted in the sense of Dirac \cite{dir}. As noted in
\cite{sm} $G$ (Poisson)commutes with $H$: $ \{G,H\}=0$ that can be
checked explicitly. So far everything appears to be a
straightforward extension of \cite{sm} but now comes the new
elements.

In two dimensions  one can moot the idea of a complex angular
momentum and demand its reality.  The  complex angular momentum is
defined as,
\begin{eqnarray}
{\cal{L}}=z_1 \pi_2 - z_2 \pi_1 \equiv L_R +i L_G , \label{1.6}
\end{eqnarray}
\begin{eqnarray}
L_R = \epsilon_{ij}(x_i p_j+ y_i q_j)~~,~~L_G = -\epsilon_{ij}(x_i
q_j-y_ip_j). \label{1.7}
\end{eqnarray}
For real values of  angular momentum  we  impose $ L_G \approx 0$.
The angular momentum  $L_R$  is a conserved quantity $
\{L_R,H\}=0$.

The two dimensional CHO has two constraints $G\approx 0,~L_G\approx 0$
(\cite{sm} had one) and so we will require a full constraint
analysis \cite{dir}, as discussed in Section II. First of all
one has to obtain the full set of linearly independent constraints
such that the constraint system is stable under time translation
In the present case this is ensured by noting,
\begin{equation}
\{G,H\}\approx 0 ,~~\{L_G,H\}\approx 0. \label{con1}
\end{equation}
Next comes the classification of the constraints. In our system,
\begin{equation}
\{L_G,G\}\approx 0. \label{con}
\end{equation}
This shows that both the constraints are FCC in nature (of the
type $F_\alpha$ mentioned in Section II).

There are two generic features that are common  in the one
dimensional model \cite{sm}
and its higher dimensional extensions studied here:\\
First one is the fact that the constraint that is generated from
the reality of angular momentum commutes with $H$. This property
remains valid in the three dimensional extension as well
 and this type of additional constraints did
not appear in one dimensional case \cite{sm}. This property might
be a particular feature of the CHO model. Remember that for the
constraint $G$ that originated from the complex Hamiltonian, one
can exploit the Cauchy-Riemann conditions to show $\{G,H\}=0$ in a
model independent way. It will be
interesting to see if our result has a deeper significance.\\
The second point is related to the degrees of freedom count.
Notice that in \cite{sm} in one dimension, one extra degree of
freedom was introduced due to complexification and it can be
removed by the single FCC $G$. This is because the additional two
variables $(y,q)$ in phase space can be removed by the FCC $G$ and
a suitable gauge choice (the so called unitary gauge). Now in two
dimensions, the extension is by two degree of freedom (four
variables $(y_i,q_i;i=1,2)$  in phase space but now there are two
FCCs $G$ and $L_G$ (along with two gauge choices) to account for
them. Hence effectively the number of degrees of freedom has not
changed in the process of complexification. This property is
preserved in three dimensions as well but in a more interesting
and non-trivial way.

A constrained  Lagrangian for the CHO is,
\begin{equation}
L=x_i\dot p_i +y_i\dot q_i -H +\lambda _1G +\lambda _2L_G,
\label{1.315}
\end{equation}
$\lambda _1,\lambda _2 $ being Lagrange multipliers. From the
Euler-Lagrange equations of motion we obtain,
\begin{equation}
p_i=\dot x_i-\lambda_1q_i-\lambda_2\epsilon_{ij}y_j,~ q_i=-\dot
y_i+\lambda_1p_i-\lambda_2\epsilon_{ij}x_j. \label{1.316}
\end{equation}
Substituting the momenta in (\ref{1.315}) and finally eliminating
the multipliers $\lambda _1,\lambda _2$  we can  get the
coordinate space Lagrangian. One can check that it is invariant
under the gauge transformations generated by $G$ and $L_G$.

In the present work we will not try to develop the full dynamics
of the model but will only
 show that the model admits closed
trajectories for positive energies and angular momentum, in a
partially gauge fixed setup  (similar to \cite{sm}) with $ \lambda
_1=\lambda _2=0$. Let us consider the simplest possible bounded
solution,
\begin{eqnarray}
x_i=A_iCos(t)+B_iSin(t);\,\,\,y_i=R_iCos(t)+Q_iSin(t)
\label{1.311}
\end{eqnarray}
where $A_i,B_i,R_i,Q_i$ are time independent parameters.
  Substituting (\ref{1.311}) in the previously computed expressions for the Hamiltonian $H$, angular
momentum $L_R$ and constraints $G~,L_G$, we obtain,
\begin{eqnarray}
&&H=\frac{1}{2}[(A^2+B^2)-(R^2+Q^2)];\,\,\,G=AR+BQ \nonumber
\\
&&L_R =
\epsilon_{ij}(A_iB_j+Q_iR_j);\,\,\,L_G=\epsilon_{ij}(A_iQ_j-B_iR_j)
\label{1.3111}
\end{eqnarray}
Now consider the following choices of $A,B,Q,R$ for which both the constraints vanish and $H$
and $L_R$ take different forms:\\
(i) $A_i=-\epsilon_{ij}B_j~;~Q_i=-\epsilon_{ij}R_j \Rightarrow H \equiv E=A^2-R^2,~L_R=-(A^2+R^2)$,\\
(ii)  $A_i=B_i=0 \Rightarrow E=-\frac{1}{2}(R^2+Q^2),~L_R=\epsilon_{ij}Q_iR_j$,\\
(iii)  $A_i=\pm\epsilon_{ij}R_j;~ B_i=\pm\epsilon_{ij}Q_j \Rightarrow E=L_R=0$,\\
(iv)  $A_i=\epsilon_{ij}B_j~;~Q_i=\epsilon_{ij}R_j \Rightarrow E= A^2-R^2, ~L_R=A^2+R^2$,\\
(v)  $R_i=Q_i=0 \Rightarrow E=\frac{1}{2}(A^2+B^2),~L_R=\epsilon_{ij}Q_iR_j$.\\
Let us now comment on these alternative possibilities: clearly the
 choices (i) and (ii) are not interesting because for classical
systems, energy or angular momentum can not be negative. Also
(iii) does not represent a dynamical system  since both energy and
angular momentum vanish. The choices (iv) and (v) are physically
relevant. Obviously (v) represents the conventional harmonic
oscillator. Let us focus our attention on (iv). Here the energy
$E$ is not positive definite but angular momentum $L_R$ is
positive definite. Turning this around, we might demand both
positive definite values for $E$ and $L_R$ and in that case we can
plot the constant (positive) energy and angular surfaces to get an
idea of the particle trajectory. Clearly a fixed positive energy
can give rise to unbounded motion in the form of open surfaces
whereas a fixed  angular momentum will lead to a closed surface,
(in fact a hyper-sphere). Hence their intersection will yield a
closed trajectory. This is shown in the figure  where both the
surfaces
are plotted with the coordinate $y_2=0$.\\
\vskip .5cm {(IV) \bf{3-Dimensional CHO: Classical Analysis:}}  The
3-dimensional CHO is studied in the same way as before where the
equations (\ref{1.1},\ref{1.5}) for the complex Hamiltonian remains structurally identical  with $i=1,2,3$.
${\cal{H}}$ will be real provided  $G\approx 0$ is treated as a
constraint.

Proceeding in the same way as we did in the 2-dimensional
counterpart in Section III we define the $i^{th}$  component
of the angular momentum as,
\begin{equation}
L^i=L_R^i+iL_G^i;~ L_R^i=\epsilon^{ijk}(x^jp^k+y^jq^k) , ~
L_G^i=\epsilon^{ijk}(y^jp^k-x^jq^k) .\label{1.36}
\end{equation}
We impose further constraints $L_G^i\approx 0$ to keep reality of
angular momenta intact. From $
\{L_R^i,H\}=0,~\{L_R^i,L_R^j\}=\epsilon^{ijk}L_R^k$ we find  $L_R$
is conserved and preserve  $SO(3)$ algebra.

Next we carry out the constraint analysis with the four
constraints, $ G\approx 0,~L_G^i \approx 0~,i=1,2,3$. From $
\{G,H\}=0,~\{L_G^i,H\}=0$, we find the system of constraints is
stable against time translation. Next   the constraint algebra $
\{L_G^i,G\}=0, ~ \{L_G^i,L_G^j\}=-\epsilon^{ijk}L_R^k$  indicates
 that $G$ is an FCC (of the type $F_\alpha $) but also there are  SCCs (of the type $H_\beta $). Since there can not be an odd number of SCCs {\footnote{Remember that the constraint matrix for SCCs is non-singular.}} (three in the
present case)  there has to be another FCC. Taking help from the
rest of the algebra $ \{L_R^i,G\}=
0,~\{L_G^i,L_R^j\}=\epsilon^{ijk}L_G^k $ we find that the
following combination, $ W\equiv L_R^iL_G^i \approx 0 $,
constitutes the other FCC. Hence we conclude that the system has
two FCCs $G\approx 0~,~W\approx 0$ (of type $F_\alpha $ in Section II) and two SCCs which we can
chosen as $L_G^1~,L_G^2 $ (of type $H_\beta $ of Section II) with the non-vanishing bracket
\begin{equation}
\{L_G^1,L_G^2\}=-L_R^3 . \label{c7}
\end{equation}

Let us consider the degrees of freedom count in presence of the
constraints. In three dimensions we have introduced three
additional degrees of freedom and they can be accounted for by the
two FCCs (each removing one degree of freedom) and the pair of SCC
(the latter together removes one degree of freedom). In this sense
the parity is once again restored between the number of degrees of
freedom in the original system and the constrained "Crypto"
system.

Although we will not pursue the quantization of the three
dimensional CHO in the present work we note that the closed
algebra of $L_R^i ,L_G^j$ is nothing but the group algebra of
$SL(2,C)$. We also stress that in the oscillator basis were
$L_R^3$ is diagonal, the SCC structure $\{L_G^1,L_G^2\}=-L_R^3$ is
{\it{not operator valued}} and so the quantization should not be
problematic. Interestingly, for the zero angular momentum state
$\{L_G^1,L_G^2\}=0$ meaning that there are no SCC for this
particular state. But even with four FCCs the degrees of freedom
still matches because remember that the zero angular momentum
state will depend only on the planar distance and not on the
angle. 
\vskip .5cm {(V) \bf{2-Dimensional CHO: Quantum Analysis:}} In
this  section we discuss the quantization of the planar CHO.
Following the procedure one adopts in the case of a normal HO, we
define two sets of lowering  operators as,
\begin{eqnarray}
a_i = \frac{1}{\surd{2}}(p_i-ix_i),\,\,\,\;~ b_i =
\frac{1}{\surd{2}}(q_i-iy_i), \label{1.8}
\end{eqnarray}
with the non-zero commutator,
$[a_i,{a_j}^{\dag}]=[b_i,{b_j}^{\dag}]=\delta_{ij}$. Next we
define the Schwinger operators ,
\begin{eqnarray}
A_1 = \frac{1}{\surd{2}}(a_1+ia_2);\,\,\,\,A_2 =
\frac{1}{\surd{2}}(a_1-ia_2) \nonumber
\\
B_1 = \frac{1}{\surd{2}}(b_1+ib_2);\,\,\,\,B_2 =
\frac{1}{\surd{2}}(b_1-ib_2) \label{1.10}
\end{eqnarray}
The only non-zero commutators are
$[A_i,{A_j}^{\dag}]=[B_i,{B_j}^{\dag}]=\delta_{ij}$. The advantage
of using $A_i,B_i$ is that both the Hamiltonian $H$ as well as the
single component of   angular momentum $L_R$  are diagonal when
expressed in terms of $A_i,B_i$. Hence we find,
\begin{equation}
H= N_{A_1}+N_{A_2}-N_{B_1}-N_{B_2},~ L_R =
N_{A_2}+N_{B_2}-N_{A_1}-N_{B_1} ,\label{1.12a}
\end{equation}
\begin{equation}
G=-(A_1B_2+A_2B_1+{A_1}^{\dag}{B_2}^{\dag}+{A_2}^{\dag}{B_1}^{\dag}),~
L_G =
A_1B_2-A_2B_1+{A_1}^{\dag}{B_2}^{\dag}-{A_2}^{\dag}{B_1}^{\dag}
\label{1.12}
\end{equation}
where the number operators are defined as
$N_{A_1}={A_1}^{\dag}A_1$ etc.. Since $H$ and $L_R$ commute, it is
possible to choose a common eigen-basis of both $H$ and $L_R$. We
choose the common eigen-basis as
$|n_{A_1},n_{B_1};n_{A_2},n_{B_2}\rangle$ with the following
action of the Schwinger operators on them:
\begin{eqnarray}
&&A_1|n_{A_1},n_{B_1};n_{A_2},n_{B_2}\rangle=\sqrt{n_{A_1}}|n_{A_1}-1,n_{B_1};n_{A_2},n_{B_2}\rangle
\nonumber
\\
&&{A_1}^{\dag}|n_{A_1},n_{B_1};n_{A_2},n_{B_2}\rangle=\sqrt{n_{A_1}+1}|n_{A_1}+1,n_{B_1};n_{A_2},n_{B_2}\rangle
\label{1.14}
\end{eqnarray}
The actions of the rest of the operators $A_2$,${A_2}^{\dag}$, $B_1$, $B_2$, ${B_1}^{\dag},
{B_2}^{\dag}$ are similar. Eigenvalues for $H$ and $L_R$ are given below:
\begin{eqnarray}
H|n_{A_1},n_{B_1};n_{A_2},n_{B_2}\rangle=(n_{A_1}+n_{A_2}-n_{B_1}-n_{B_2})|n_{A_1},n_{B_1};n_{A_2},n_{B_2}\rangle
\nonumber
\\
L_R|n_{A_1},n_{B_1};n_{A_2},n_{B_2}\rangle=(n_{A_2}+n_{B_2}-n_{A_1}-n_{B_1})|n_{A_1},n_{B_1};n_{A_2},n_{B_2}\rangle .
\label{1.15}
\end{eqnarray}
Any state can be written as a linear combination in the above
basis,
\begin{eqnarray}
|\Psi\rangle = \sum_{n_{A_1},....,n_{B_2}=0}^{\infty}
C_{n_{A_1},....,n_{B_2}} |n_{A_1},n_{B_1};n_{A_2},n_{B_2}\rangle
\label{1.16}
\end{eqnarray}
Now comes the role of the constraints. Since they are FCCs we
follow the Dirac formalism \cite{dir} and pick the physical sector by
demanding that the FCCs kill the physical states
$(FCC)|\Psi^{ph}\rangle = 0$ which in the present case means:
\begin{eqnarray}
G|\Psi^{ph}\rangle = 0;\,\,\,\,\, L_G|\Psi^{ph}\rangle = 0.
\label{1.17}
\end{eqnarray}
However, in the present problem, it is more convenient to impose the linear combinations of
FCCs,
\begin{eqnarray}
(G+L_G)|\Psi^{ph}\rangle = 0;\,\,\,\,\, (G-L_G)|\Psi^{ph}\rangle =
0. \label{01.17}
\end{eqnarray}
Considering the first one $(G+L_G)|\Psi^{ph}\rangle = 0$,  we find,
\begin{eqnarray}
&& \sum_{n_{A_1},....,n_{B_2}=0}^{\infty}
C_{n_{A_1},....,n_{B_2}}[\sqrt{n_{A_2}n_{B_1}}
|n_{A_1},n_{B_1}-1;n_{A_2}-1,n_{B_2}\rangle \nonumber
\\
&&+\sqrt{(n_{A_2}+1)(n_{B_1}+1)}
|n_{A_1},n_{B_1}+1;n_{A_2}+1,n_{B_2}\rangle]=0 \label{1.18}
\end{eqnarray}
To find the states that satisfy (\ref{1.18}) with arbitrary  energy $m$ (including zero) and
arbitrary values of angular momentum $n$ (including zero), we use
(\ref{1.15}) and obtain the conditions,
\begin{equation}
n_{A_1}+n_{A_2}=m+ n_{B_1}+n_{B_2},~
n_{A_2}-n_{A_1}=n+n_{B_1}-n_{B_2}, \label{1.20}
\end{equation}
where $m = -\infty$ to $+\infty$ and $n=0,1,2,3,.........$. The
numbers $n_A,n_B$ are the eigen-values of the corresponding number
operators $N_A,N_B$ etc.. Solving the above two equations we get,
\begin{equation}
n_{B_1}=n_{A_2}-\frac{m}{2}-\frac{n}{2},~
n_{B_2}=n_{A_1}-\frac{m}{2}+\frac{n}{2}. \label{1.21}
\end{equation}
Substituting these in (\ref{1.18}) we have,
\begin{eqnarray}
\sum_{n_{A_1},n_{A_2}=0}^{\infty}
C_{n_{A_1},n_{A_2}}[\sqrt{(n_{A_2}-\frac{m}{2}-\frac{n}{2})n_{A_2}}
&&|n_{A_1},n_{A_2}-\frac{m}{2}-\frac{n}{2}-1; \nonumber
\\
&&n_{A_2}-1,n_{A_1}+\frac{n}{2}-\frac{m}{2}\rangle \nonumber
\\
+~\sqrt{(n_{A_2}+1)(n_{A_2}-\frac{m}{2}-\frac{n}{2}+1)}
&&|n_{A_1},n_{A_2}-\frac{m}{2}-\frac{n}{2}+1; \nonumber
\\
&&n_{A_2}+1,n_{A_1}+\frac{n}{2}-\frac{m}{2}\rangle] \nonumber
\\
&&=0 \label{1.22}
\end{eqnarray}
Replacing $n_{A_1}$ by $n_1$ and $n_{A_2}$ by $n_{2}$ and then
substituting $n_2$ by $n_2 -2$ in the second term of the above
relation we get,
\begin{eqnarray}
\sum_{n_{1},n_{2}=0}^{\infty} C_{n_{1},n_{2}}
\sqrt{(n_{2}-\frac{m}{2}-\frac{n}{2})n_{2}}
&&|n_{1},n_{2}-\frac{m}{2}-\frac{n}{2}-1; \nonumber
\\
&&n_{2}-1,n_{1}-\frac{m}{2}+\frac{n}{2}\rangle \nonumber
\\
+\sum_{n_{1}=0,n_{2}=2}^{\infty}
C_{n_{1},n_{2}-2}\sqrt{(n_{2}-\frac{m}{2}-\frac{n}{2}-1)(n_{2}-1)}
&&|n_{1},n_{2}-\frac{m}{2}-\frac{n}{2}-1; \nonumber
\\
&&n_{2}-1,n_{1}-\frac{m}{2}+\frac{n}{2}\rangle] \nonumber
\\
&&=0 \label{1.23}
\end{eqnarray}
Putting $C_{n_1,-2}=C_{n_1,-1}=0$ we can rewrite the above as,
\begin{eqnarray}
&&\sum_{n_{1},n_{2}=0}^{\infty} [C_{n_{1},n_{2}}
\sqrt{(n_{2}-\frac{m}{2}-\frac{n}{2})n_{2}}~
+~C_{n_{1},n_{2}-2}\sqrt{(n_{2}-\frac{m}{2}-\frac{n}{2}-1)(n_{2}-1)}]
\nonumber
\\
&&|n_{1},n_{2}-\frac{m}{2}-\frac{n}{2}-1;n_{2}-1,n_{1}-\frac{m}{2}+\frac{n}{2}\rangle
\nonumber
\\
&&=0
 \label{1.24}
\end{eqnarray}
Since the basis vectors are linearly independent, the coefficient
within the third bracket must vanish for each basis vectors
 and hence we have the following recursion relation,
\begin{eqnarray}
C_{n_{1},n_{2}}
=-\sqrt{\frac{(n_{2}-\frac{m}{2}-\frac{n}{2}-1)(n_{2}-1)}{(n_{2}-\frac{m}{2}-\frac{n}{2})n_{2}}}
C_{n_{1},n_{2}-2} \label{1.25}
\end{eqnarray}
where $n_1=0,1,2,3,..........$ and $n_2=2,3,4,.......$ since
$C_{n_1,-2}=C_{n_1,-1}=0$. Using the above recursion relation one
can show,
\begin{eqnarray}
C_{i,2k}=(-1)^k
\sqrt{\frac{(2k-1-\frac{m}{2}-\frac{n}{2})!!(2k-1)!!}{(2k-\frac{m}{2}-\frac{n}{2})!!(2k)!!}}
C_{i,0}~; \nonumber
\\
C_{i,2k+1}=(-1)^k
\sqrt{\frac{(2k-\frac{m}{2}-\frac{n}{2})!!(2k)!!}{(2k+1-\frac{m}{2}-\frac{n}{2})!!(2k+1)!!}}
C_{i,1} \label{1.26}
\end{eqnarray}
where $i=0,1,2,........$, $k=1,2,3,......$ and
\begin{eqnarray}
(2k)!!&=&2.4.6......(2k-2)(2k); \nonumber
\\
(2k-1)!!&=&1.3.5....(2k-3)(2k-1); \nonumber
\\
(2k-1-\frac{m}{2}-\frac{n}{2})!!&=&(1-\frac{m}{2}-\frac{n}{2})(3-\frac{m}{2}-\frac{n}{2})........
\nonumber
\\
&&(2k-3-\frac{m}{2}-\frac{n}{2})(2k-1-\frac{m}{2}-\frac{n}{2});
\label{1.27}
\end{eqnarray}
Therefore by imposing one of the FCCs  we narrow down the
physical sector to the following state,
\begin{eqnarray}
|\Psi^{ph}_{{m},{n}}\rangle &&= \sum_{{i=0},{k=1}}^{\infty} (-1)^k
[\sqrt{\frac{(2k-1-\frac{m}{2}-\frac{n}{2})!!(2k-1)!!}{(2k-\frac{m}{2}-\frac{n}{2})!!(2k)!!}}
\nonumber
\\
&&C_{i,0}|i,2k-\frac{m}{2}-\frac{n}{2};2k,i-\frac{m}{2}+\frac{n}{2}\rangle
\nonumber
\\
&&+~\sqrt{\frac{(2k-\frac{m}{2}-\frac{n}{2})!!(2k)!!}{(2k+1-\frac{m}{2}-\frac{n}{2})!!(2k+1)!!}}
\nonumber
\\
&&C_{i,1}|i,2k+1-\frac{m}{2}-\frac{n}{2};2k+1,i-\frac{m}{2}+\frac{n}{2}\rangle]
\label{1.28}
\end{eqnarray}
Finally we further restrict the sector to the correct physical one
by imposing the other FCC,
\begin{equation}
(G-L_G)|\Psi^{ph}_{{m},{n}}\rangle=
-2(A_1B_2+A_{1}^{\dag}B_{2}^{\dag})|\Psi^{ph}_{{m},{n}}\rangle =
0. \label{ph1}
\end{equation}
 Another relation between the parameters follows:
\begin{eqnarray}
&&\sum_{{i=0},{k=1}}^{\infty} (-1)^k
[\sqrt{\frac{(2k-1-\frac{m}{2}-\frac{n}{2})!!(2k-1)!!}{(2k-\frac{m}{2}-\frac{n}{2})!!(2k)!!}}
\nonumber
\\
&&C_{i,0}\{\sqrt{i(i-\frac{m}{2}+\frac{n}{2})}|i-1,2k-\frac{m}{2}-\frac{n}{2};2k,i-\frac{m}{2}+\frac{n}{2}-1\rangle
\nonumber
\\
&&+~\sqrt{(i+1)(i+1-\frac{m}{2}+\frac{n}{2})}|i+1,2k-\frac{m}{2}-\frac{n}{2};2k,i-\frac{m}{2}+\frac{n}{2}+1\rangle\}
\nonumber
\\
&&+~\sqrt{\frac{(2k-\frac{m}{2}-\frac{n}{2})!!(2k)!!}{(2k+1-\frac{m}{2}-\frac{n}{2})!!(2k+1)!!}}
\nonumber
\\
&&C_{i,1}\{\sqrt{i(i-\frac{m}{2}+\frac{n}{2})}|i-1,2k+1-\frac{m}{2}-\frac{n}{2};2k+1,i-\frac{m}{2}+\frac{n}{2}-1\rangle
\nonumber
\\
&&+\sqrt{(i+1)(i-\frac{m}{2}+\frac{n}{2}+1)}|i+1,2k+1-\frac{m}{2}-\frac{n}{2};2k+1,i-\frac{m}{2}+\frac{n}{2}+1\rangle\}]
\nonumber
\\
&&=0 \label{1.29}
\end{eqnarray}
Explicitly writing the above equation for the sum over $i=0$ to
$\infty$ one can show,
\begin{eqnarray}
&&C_{\mu\nu}=0; \nonumber
\\
&&C_{2r,0}=(-1)^r\sqrt{\frac{(2r-1)!!(2r-1-\frac{m}{2}+\frac{n}{2})!!}{(2r)!!(2r-\frac{m}{2}+\frac{n}{2})!!}}
C_{0,0} \nonumber
\\
&&C_{2r,1}=(-1)^r\sqrt{\frac{(2r-1)!!(2r-1-\frac{m}{2}+\frac{n}{2})!!}{(2r)!!(2r-\frac{m}{2}+\frac{n}{2})!!}}
C_{0,1} \label{1.30}
\end{eqnarray}
where $\mu=1,3,5,7,........$; $\nu=0,1$ and $r=
1,2,3,4,..........$. So the final form of the physical state  for
arbitrary energy $m$ and angular momentum $n$ is,
\begin{eqnarray}
&&|\Psi^{ph}_{{m},{n}}\rangle = \sum_{{r,k=1}}^{\infty} (-1)^{k+r}
\sqrt{\frac{(2r-1)!!(2r-1-\frac{m}{2}+\frac{n}{2})!!}{(2r)!!(2r-\frac{m}{2}+\frac{n}{2})!!}}
\nonumber
\\
&&[\sqrt{\frac{(2k-1-\frac{m}{2}-\frac{n}{2})!!(2k-1)!!}{(2k-\frac{m}{2}-\frac{n}{2})!!(2k)!!}}
C_{0,0}|2r,2k-\frac{m}{2}-\frac{n}{2};2k,2r-\frac{m}{2}+\frac{n}{2}\rangle
\nonumber
\\
&&+~\sqrt{\frac{(2k-\frac{m}{2}-\frac{n}{2})!!(2k)!!}{(2k+1-\frac{m}{2}-\frac{n}{2})!!(2k+1)!!}}
C_{0,1} \nonumber
\\
&&|2r,2k+1-\frac{m}{2}-\frac{n}{2};2k+1,2r-\frac{m}{2}+\frac{n}{2}\rangle
] \label{1.31}
\end{eqnarray}
With this we conclude the quantization of the 2-dimensional
Crypto-oscillator.

 It is also straightforward to recover the quantum version of 1-dimensional
CHO, that was discussed in  \cite{sm}. In one dimension,
$x_2,p_2,y_2,q_2$ are absent from the set (\ref{1.8}) which means
that in (\ref{1.10}) $A_1=A_2\equiv A,~B_1=B_2\equiv B$. Putting
this back in (\ref{1.12},\ref{1.12}), we obtain,
\begin{equation}
H=2(N_A-N_B),~~G=-2(AB+A^{\dag}B^{\dag}),~~L_R=0,~~L_G=0,
\label{s}
\end{equation}
which is nothing but the model studied in \cite{sm}. \vskip .5cm
{(VI) \bf{Summary and Outlook:}} In this paper we have generalized the
Crypto Harmonic Oscillator model, proposed by Smilga \cite{sm}, to
higher (two and three) dimensions. After complexification, the
energy is restricted to the  real sector by demanding that the
imaginary part of the energy vanish. This introduces a
(Hamiltonian) constraint in the theory \cite{sm}. In higher
dimensions there are other physical dynamical variables (such as
angular momentum that is considered here) besides the energy and
it is only natural to restrict them to the real sector as well.
This brings in additional constraints and a formal constraint
analysis \cite{dir} reveals interesting features. Also we have
quantized the two dimensional Crypto Harmonic Oscillator in the
present paper.

An interesting  problem  is to ascertain to what extent the new
features in the constraint structure revealed here in the higher
dimensional extension, are model independent. If these features
turn out to be  generic,  then this formalism can be still another
alternative way of introducing gauge symmetry via
 phase space extension. In fact we are now studying the Crypto version
 of the oscillator with a position dependent effective mass and there also
 these features persist. These results will be reported elsewhere.

The other problem is obviously to apply this idea of Crypto-gauge
invariance, as adapted in our work in higher space dimensions, to
more complicated models and to compare the results with the
analogue higher dimensional $PT$-symmetric models.

\vskip .2cm \noindent {\it{Acknowledgements}}: It is a pleasure to
thank Rabin Banerjee and Pinaki Roy for discussions. 
\end{document}